\documentstyle[12pt]{article}
\advance\hoffset by -7mm
 \makeatletter
\@addtoreset{equation}{section} \makeatother

\renewcommand{\title}[1]{\null\vspace{25mm}

\noindent{\Large{\bf #1}}\vspace{10mm} }
\newcommand{\authors}[1]{\noindent{\large #1}\vspace{3mm}

}
\newcommand{\address}[1]{\noindent #1\vspace{5mm}

}
\renewcommand{\abstract}[1]{\vspace{19mm}

\noindent{\small{\em Abstract.} #1}\vspace{2mm}
\newcommand\bea{\begin{eqnarray}}
\newcommand\eea{\end{eqnarray}}
}
\begin{document}
\title{Density Operator and Entropy \\[2mm]
of the Damped Quantum Harmonic Oscillator
}

\authors{A. Isar}
\address{Department of Theoretical Physics, Institute of Atomic Physics\\
Bucharest-Magurele, POB MG-6, Romania\\
e-mail: isar@theory.nipne.ro}
\abstract{ The expression for the
density operator of the damped harmonic oscillator is derived from
the master equation in the framework of the Lindblad theory for open
quantum systems. Then the von Neumann entropy and effective
temperature of the system are obtained. The entropy for a state
characterized by a Wigner distribution function which is Gaussian in
form is found to depend only on the variance of the distribution
function. }

\section{Introduction}

In the last two decades, the problem of dissipation in quantum mechanics, i.e.
the consistent description of open quantum systems, was investigated by various
authors [1-5]  (for a recent review see ref. \cite{rev}).
It is commonly understood \cite{d2,d1} that dissipation
in an open system results from microscopic reversible interactions between
the observable system and the environment.
Because dissipative processes imply irreversibility and,
therefore, a preferred direction in time, it is generally thought that quantum
dynamical semigroups are the basic tools to introduce dissipation in quantum
mechanics. In the Markov approximation
the most general form of the generators of such semigroups was given
by Lindblad \cite{l1}. This formalism has been studied for the case of damped
harmonic oscillators \cite{l2,ss} and applied to various physical phenomena,
for instance, the damping of collective modes in deep inelastic collisions in
nuclear physics \cite{i1}. In \cite{i2} the Lindblad master
equation for the harmonic oscillator
was transformed into Fokker-Planck equations for quasiprobability distributions
and a comparative study was made for the Glauber $P$, antinormal ordering $Q$
and Wigner $W$ representations. In \cite{a} the density matrix for the coherent
state representation and the Wigner distribution function subject to different
types of initial conditions were obtained for the damped harmonic oscillator.
From a preceding analysis \cite{ass}
resulted an analogy in form of the Lindblad master and Fokker-Planck equations
with the corresponding equations from quantum optics, which is
based on the quantum mechanics
of the damped harmonic oscillator \cite{d2} and the corresponding
Brownian motion.
The equation of motion generally used in the theory of Brownian
motion is the master equation for the density operator or the
Fokker-Planck equation satisfied by the distribution function \cite{d2,hh,hr}.

In the present study we are also concerned
with the observable system of a harmonic oscillator which interacts with
an environment.
The aim of this work is to explore further the physical aspects of the
Fokker-Planck
equation which is the $c$-number equivalent equation to the master
equation. Generally the master equation for the density operator gains
considerably in clarity if it is represented in terms of the Wigner
distribution function which satisfies the Fokker-Planck equation.
Within the Lindblad theory for open quantum systems, we will describe
the evolution of the considered system towards a final equilibrium state.
First we derive a closed form of the density operator
satisfying the Lindblad master equation
and then calculate the von Neumann entropy and the effective temperature
of the quantum-mechanical system in a state characterized by a Wigner
distribution function which is Gaussian in form. The two quantities
are shown to evolve to their equilibrium values.

Physically, entropy can be interpreted as a measure
of the lack of knowledge (disorder) of the system. The effective
temperature associated with the Bose distribution has
been introduced \cite{lw} in connection with the entropy
obtained in the quantum theory of relaxation for the harmonic oscillator.
The entropy for an infinite
coupled harmonic oscillator chain has also been calculated for
classical \cite{rh} and quantum mechanical systems \cite{a1} represented by a
phase space distribution function. In the present work we extract the density
operator of the damped harmonic oscillator in the Lindblad theory by
using a technique analogous to those applied in the description
of quantum relaxation \cite{lw,a1,m,j1}.
Denoting by $ \rho(t)$ the density operator of the damped harmonic
oscillator in the Schr\"odinger
picture, the von Neumann entropy  $S(t) $ is given by the expectation value
of the logarithmic operator $\ln\rho:$
\bea   S(t)=-k<\ln\rho>=-k{\rm Tr}(\rho\ln\rho),     \eea
where $k$ is Boltzmann's constant.
Accordingly, the calculation of the entropy reduces to the
problem of finding the explicit form of the density operator.

The content of this paper is arranged as follows. In Sec. 2 we write
the master equation for the density operator of the harmonic
oscillator. In Sec. 3 we derive a closed form of the density operator
satisfying the master equation based on the Lindblad dynamics.
By using the explicit form of the density operator,
in Sec. 4 we calculate the von Neumann entropy and time-dependent
temperature  and analyze their temporal behaviour.
Finally, concluding remarks are given in Sec. 5.

\section{Master equation for the damped quantum harmonic oscillator}

The rigorous description of the dissipation in a quantum
mechanical system is based on the quantum dynamical semigroups \cite{d,s,l1}.
According to
the axiomatic theory of Lindblad \cite{l1}, the usual von Neumann-Liouville
equation ruling the time evolution of closed quantum systems is replaced in the
case of open systems by the following equation for the density operator $\rho$:
\bea   {d\Phi_{t}(\rho)\over dt}=L(\Phi_{t}(\rho)).     \eea
Here, $\Phi_{t}$ denotes the dynamical semigroup describing the irreversible
time evolution of the open system in the Schr\"odinger representation and $L$
the infinitesimal generator of the dynamical semigroup $\Phi_t$. Using the
structural theorem of Lindblad \cite{l1}, which gives the most general form of
the
bounded, completely dissipative Liouville operator $L$, we obtain the explicit
form of the most general time-homogeneous quantum mechanical Markovian master
equation:
\bea   {d\rho(t)\over dt}=L(\rho(t)),     \eea
where
\bea   L(\rho(t))=-{i\over\hbar}[H,\rho(t)]+{1\over 2\hbar}\sum_{j}([V_{j}
\rho(t),V_{j}^\dagger]+[V_{j},\rho(t)V_{j}^\dagger]).     \eea
Here $H$ is the Hamiltonian of the system. The operators $V_{j}$ and $V_{j}^
\dagger$
are bounded operators on the Hilbert space $\cal H$ of the Hamiltonian.

We should like to mention that the Markovian master equations found in the
literature are of this form after some rearrangement of terms, even for
unbounded Liouville operators. In this connection we assume that the general
form of the master equation given by (2.2), (2.3) is also valid for unbounded
Liouville operators.

In this paper we impose a simple condition to the operators $H,V_{j},V_{j}^
\dagger$
that they are functions of the basic observables $\hat q$ and $\hat p$ of the
one-dimensional quantum mechanical system (with $[\hat q,\hat p]=i\hbar
I,$ where $I$ is the identity operator on $\cal H$) of such kind
that the obtained model is exactly solvable. A precise version for this last
condition is that linear spaces spanned by first degree (respectively second
degree) noncommutative polynomials in $\hat q$ and $\hat p $ are invariant to
the action
of the completely dissipative mapping $L$. This condition implies \cite{l2}
that $V_{j}$
are at most first degree polynomials in $\hat q$ and $\hat p $ and $H$ is
at most a second degree polynomial in $\hat q$ and $\hat p $. Then the harmonic
oscillator Hamiltonian $H$ is chosen of the general form
\bea   H=H_{0}+{\mu \over 2}(\hat q\hat p+\hat p\hat q),~~~H_{0}={1\over 2m}
\hat p^2+{m\omega^2\over 2}\hat q^2.     \eea
With these choices the Markovian master equation can be written \cite{rev,ss}:
\bea   {d\rho \over dt}=-{i\over \hbar}[H_{0},\rho]-{i\over 2\hbar}(\lambda +
\mu)[\hat q,\rho \hat p+\hat p\rho]+{i\over 2\hbar}(\lambda -\mu)[\hat p,\rho
\hat q+\hat q\rho]  \nonumber\\
  -{D_{pp}\over {\hbar}^2}[\hat q,[\hat q,\rho]]-{D_{qq}\over {\hbar}^2}
[\hat p,[\hat p,\rho]]+{D_{pq}\over {\hbar}^2}([\hat q,[\hat p,\rho]]+[\hat p,
[\hat q,\rho]]),      \eea
where $D_{pp},D_{qq}$ and $D_{pq}$ are the diffusion coefficients and $\lambda$
the friction constant. They satisfy the following fundamental constraints
\cite{rev,ss}:
\bea   {\rm i})~D_{pp}>0,~~{\rm ii})~D_{qq}>0,~~{\rm iii})~D_{pp}D_{qq}-D_{pq}
^2\ge{\lambda}^2{\hbar}^2/4.    \eea
In the particular case when the asymptotic state is a Gibbs state
\bea   \rho_G(\infty)=e^{-{H_0\over kT}}/{\rm Tr}e^{-{H_0\over kT}},    \eea
these coefficients reduce to
\bea   D_{pp}={\lambda+\mu\over 2}\hbar m\omega\coth{\hbar\omega\over 2kT},
~~D_{qq}={\lambda-\mu\over 2}{\hbar\over m\omega}\coth{\hbar\omega\over 2kT},
~~D_{pq}=0,    \eea
where $T$ is the temperature of the thermal bath and the
fundamental constraints are satisfied only if $\lambda>|\mu|.$

\section{Density operator of the damped harmonic oscillator}

By introducing the real variables $x_1, x_2$ corresponding to the
operators $\hat q, \hat p$:
\bea x_1=\sqrt{m\omega\over 2\hbar}q, ~~x_2={1\over\sqrt{2\hbar m\omega}}p,\eea
in \cite{i2,a} we have transformed the master
equation for the density operator into the following Fokker-Planck equation
satisfied by the Wigner distribution function $W(x_1, x_2, t):$
\bea   {\partial W\over\partial t}=\sum_{i,j=1,2}A_{ij}{\partial\over
\partial x_i}(x_jW)+{1\over 2}\sum_{i,j=1,2}Q^W_{ij}{\partial^2\over
\partial x_i\partial x_j}W,    \eea
where
\bea   A=\left(\matrix{\lambda-\mu&-\omega\cr
\omega&\lambda+\mu\cr}\right),~~~
Q^W={1\over\hbar}\left(\matrix{m\omega D_{qq}&D_{pq}\cr
D_{pq}&D_{pp}/m\omega\cr}\right).    \eea
Since the drift coefficients are linear in the variables $x_1$ and $x_2$ and
the diffusion coefficients are constant with respect to $x_1$ and $x_2,$
Eq. (3.2)
describes an Ornstein-Uhlenbeck process \cite{wu}. Following the method
developed by Wang and Uhlenbeck \cite{wu}, we solved in ref. \cite{a}
this Fokker-Planck equation, subject
to either the wave-packet type or the $\delta$-function type of initial
conditions.

In the present paper we consider the underdamped case
($|\mu|<\omega$) of the harmonic oscillator \cite{rev,ss}.
If the initial condition for the Fokker-Planck equation is of a Gaussian
(wave-packet) type ($x_{10}$ and $x_{20}$ are the
initial values of $x_1$ and $x_2$ at $t=0,$ respectively)
\bea W_w(x_1, x_2, 0)={1\over\pi\hbar}\exp\{-2[(x_1-x_{10})^2+(x_2-x_{20})^2
]\},  \eea
the solution of Eq. (3.2) is given by \cite{a}:
\bea W_w(x_1, x_2)={\Omega\over \pi\hbar\omega\sqrt{-B_w}}\exp\{-{1\over B_w}
[\phi_w(x_1-\bar x_1)^2+\psi_w(x_2-\bar x_2)^2+\chi_w(x_1-\bar x_1)
(x_2-\bar x_2)]\},
\eea
where
\bea   B_w=g_1g_2-{1\over 4}g_3^2,~~g_1=g_2^*={\mu a\over\omega}e^{2\Lambda t}+
{d_1\over \Lambda}(e^
{2\Lambda t}-1),~~g_3=2[e^{-2\lambda t}+{d_2\over\lambda}(1-e^{-2\lambda t})],
    \eea
\bea\phi_w=g_1a^{*2}+g_2a^2-g_3,~\psi_w=g_1+g_2-g_3,~\chi_w=2(g_1a^*+g_
2a)-g_3(a+a^*).  \eea
We have put $a=(\mu-i\Omega)/\omega, ~\Lambda=-\lambda-i\Omega,$
~$d_1=(a^2m\omega D_{qq}+2aD_{pq}+D_{pp}/m\omega)/\hbar,$
~$d_2=(m\omega D_{qq}+2\mu D_{pq}/\omega+D_{pp}/m\omega)/\hbar$
and $\Omega^2=\omega^2-\mu^2.$
The functions $\bar x_1$ and $\bar x_2$, which are also oscillating
functions, are found to be \cite{a}:
\bea\bar x_1=e^{-\lambda t}[(x_{10}(\cos\Omega t+{\mu\over\Omega}\sin
\Omega t)+x_{20}{\omega\over\Omega}\sin\Omega t],  \eea
\bea\bar x_2=e^{-\lambda t}[(x_{20}(\cos\Omega t-{\mu\over\Omega}\sin
\Omega t)-x_{10}{\omega\over\Omega}\sin\Omega t].  \eea
The solution (3.5)
of the Fokker-Planck equation (3.2), subject to the wave-packet type of initial
condition (3.4) can be written in terms of the coordinate and momentum
($<\hat A>={\rm Tr}(\rho \hat A)$ denotes the expectation value of an
operator $\hat A$) as \cite{rev,ss}:
\bea   W(q,p)={1\over 2\pi\sqrt{\delta}}\exp\{-{1\over 2\delta}[\phi
(q-<\hat q>)^2+\psi(p-<\hat p>)^2-2\chi(q-<\hat q>)(p-<\hat p>)]\},
\eea
where
\bea <\hat q>=e^{-\lambda t}[(\cos\Omega t+{\mu\over\Omega}\sin\Omega t)
<\hat q(0)>+{1\over m\Omega}\sin\Omega t<\hat p(0)>],\eea
\bea <\hat p>=e^{-\lambda t}[-{m\omega^2\over\Omega}\sin\Omega t<\hat q(0)>+
(\cos\Omega t-{\mu\over\Omega}\sin\Omega t)<\hat p(0)>],    \eea
\bea  \phi\equiv\sigma_{pp}=<\hat p ^2>-<\hat p >^2=
-{\hbar m\omega^3\over 4\Omega^2}\phi_w,    \eea
\bea  \psi\equiv\sigma_{qq}=<\hat q ^2>-<\hat q >^2
=-{\hbar\omega\over 4m\Omega^2}\psi_w, \eea
\bea  \chi\equiv\sigma_{pq}={1\over 2}<\hat q \hat p +\hat p \hat q >-
<\hat q ><\hat p >
={\hbar\omega^2\over 8\Omega^2}\chi_w,~~\delta=\phi\psi-\chi^2.\eea

To obtain the explicit form of the density operator, we apply, like
in \cite{a1,j1}, the relation $\rho=2\pi\hbar {\bf\it
N}\{W_s(q,p)\},$ where $W_s$ is the Wigner distribution function in
the form of standard rule of association and ${\bf\it N}$ is the
normal ordering operator \cite{w,l} which acting on the function
$W_s(q,p)$ moves $p$ to the right of $q.$ By the standard rule of
association we mean the correspondence $p^mq^n\to \hat q^n\hat p^m$
between functions of two classical variables $(q,p)$ and functions
of two quantum canonical operators $(\hat q,\hat p).$ The Wigner
distribution function (3.10), which is in the form of the Weyl rule
of association \cite{hw}, can be transformed into the form of
standard rule of association \cite{clm} by performing the operation:
\bea W_s(q,p)=\exp({1\over 2}i\hbar{\partial^2\over\partial
p\partial q}) W(q,p). \eea We get the following Wigner distribution
function: \bea W_s(q,p)={1\over 2\pi\sqrt{\xi}}\exp\{-{1\over
2\xi}[\phi (q-<\hat q>)^2+\psi(p-<\hat p>)^2\\-2(\chi-i{\hbar\over
2}) (q-<\hat q>)(p-<\hat p>)]\},    \eea where \bea
\xi=\phi\psi-(\chi-i{\hbar\over 2})^2.\eea The normal ordering
operator $N$ can be applied upon the Wigner function $W_s$ in
Gaussian form by using McCoy's theorem \cite{w,l}: \bea  {\bf\it
N}[\exp(Aq^2+Bp^2+Gqp)]=
 [J\exp(-i\hbar\gamma)]^{1/2}\exp(\alpha\hat q^2+\beta\hat p^2+\gamma
\hat q\hat p),\eea
where $\alpha=A/C,~ \beta=B/C,~ C=\sinh\Gamma/\Gamma J, ~\Gamma=-i\hbar(\gamma
^2-4\alpha\beta)^{1/2},$ with $J=\cosh\Gamma+i\hbar\gamma\sinh\Gamma/\Gamma
=1/(1-i\hbar G).$
After a straightforward, but lengthy calculation, we obtain the
following expression of the density operator:
\bea  \rho={\hbar\over \sqrt{\xi}}\exp\{{1\over 2}\ln{4\xi\over 4\delta-
\hbar^2}-{1\over 2\hbar\sqrt{\delta}}\cosh^{-1}(1+{2\hbar^2
\over 4\delta-\hbar^2})\nonumber\\
  \times  [ \phi(\hat q-<\hat q>)^2+\psi(\hat p-<\hat p>)^2-\chi
[2(\hat q-<\hat q>)(\hat p-<\hat p>)-i\hbar] ] \}.
      \eea
This form is analogous to those obtained in \cite{a1,j1}.
In particular, if $D_{qq}=D_{pq}=0$ and $\mu=\lambda,$ we
obtain the Jang's model \cite{j1,j2} on nuclear dynamics based
on the second RPA at finite temperature.
The density operator (3.20) has a Gaussian form, as expected from the
initial form of the Wigner distribution function. While the Wigner distribution
is expressed in terms of real variables $q$ and $p,$ the density operator is
a function of operators $\hat q$ and $\hat p.$ When time $t\to\infty,$
the density operator tends to
\bea  \rho(\infty)={2\hbar\over{\sqrt{4\sigma-\hbar^2}}}\exp\{-{1\over
2\hbar\sqrt{\sigma}}\ln{2\sqrt{\sigma}+\hbar\over 2\sqrt{\sigma}-\hbar}
[\sigma_{pp}(\infty)\hat q^2+\sigma_{qq}(\infty)\hat p^2-\sigma_{pq}(\infty)
(\hat q\hat p+\hat p\hat q)]\},    \eea
where $\sigma=\sigma_{pp}(\infty)\sigma_{qq}(\infty)-\sigma^2_{pq}(\infty)$
and \cite{rev,ss}:
\bea   \sigma_{qq}(\infty)={1\over 2m^2\omega^2\lambda(\lambda^2+\Omega^2)}
[m^2\omega^2(2\lambda(\lambda+\mu)+\omega^2)D_{qq}
  +\omega^2D_{pp}+2m\omega^2(\lambda+\mu)D_{pq}], \eea
\bea   \sigma_{pp}(\infty)={1\over 2\lambda(\lambda^2+\Omega^2)}[m^2\omega^4
D_{qq}+(2\lambda(\lambda-\mu)+\omega^2)D_{pp}-2m\omega^2(\lambda-
\mu)D_{pq}],    \eea
\bea   \sigma_{pq}(\infty)={1\over 2m\lambda(\lambda^2+\Omega^2)}[-
m^2\omega^2(\lambda+\mu)D_{qq}+(\lambda-\mu)D_{pp}+2m(\lambda^2-\mu^2)D_{pq}].
\eea
In the particular case (2.8),
\bea  \sigma_{qq}(\infty)={\hbar\over 2m\omega}\coth{\hbar\omega\over 2kT},~
\sigma_{pp}(\infty)={\hbar m\omega\over 2}\coth{\hbar\omega\over 2kT},~
 \sigma_{pq}(\infty)=0    \eea
and the asymptotic state is a Gibbs state (2.7):
\bea  \rho_G(\infty)=2\sinh{\hbar\omega\over 2kT}\exp\{-{1\over kT}({1
\over 2m}\hat p^2+{m\omega^2\over 2}\hat q^2)\}.\eea

\section{Von Neumann entropy and effective temperature}

By using the relations (3.13-15), we get the expectation value of $\ln\rho$:
\bea  <\ln\rho>=\ln\hbar-{1\over 2}\ln(\delta-{\hbar^2\over 4})-{\sqrt{\delta}
\over\hbar}\ln{2\sqrt{\delta}+\hbar\over 2\sqrt{\delta}-\hbar}.\eea
Denoting $\hbar\nu=\sqrt{\delta}-\hbar/2,$
we finally obtain the following expression of the von Neumann entropy:
\bea  S(t)=k[(\nu+1)\ln(\nu+1)-\nu\ln\nu].    \eea
Since $\delta=-{\displaystyle{\hbar^2\omega^2\over 4\Omega^2}}B_w,$
the function $\nu$ becomes
\bea  \nu={\omega\over 2\Omega}\sqrt{-B_w}-{1\over 2},\eea
where
\bea B_w=\exp(-4\lambda t)(2{\mu\over\omega}{\rm Re}{d_1 a^*\over\Lambda}
-{\Omega^2\over\omega^2}+{\vert d_1\vert^2\over\vert\Lambda\vert^2}-
{d_2^2\over\lambda^2}+2{d_2\over\lambda})\nonumber\\
-2\exp(-2\lambda t) [{\rm Re}
\left ( ({\mu\over\omega}{ d_1 a^*\over\Lambda}+{\vert d_1\vert^2\over
\vert\Lambda\vert^2})\exp 2i\Omega t \right )-
{d_2^2\over\lambda^2}+{d_2\over\lambda} ]+{\vert d_1\vert^2\over
\vert\Lambda\vert^2}-{d_2^2\over\lambda^2}.\eea
It is worth noting that the entropy depends only upon the variance of the
Wigner distribution. When time $t\to\infty,$  $\nu$ tends to $s=\omega(d_2^2/
\lambda^2-\vert d_1\vert^2/(\lambda^2+\Omega^2))^{1/2}/2\Omega-1/2$
and the entropy relaxes to its equilibrium value $S(\infty)=k[(s+1)\ln(s+1)-s
\ln s].$

The expression (4.2) is analogous to those previously obtained \cite{lw,rh}
in the theory of quantum oscillator relaxation
and for the description of a system of collective RPA phonons \cite{j1}.
It should also be noted that the expression (4.2) has the same form as the
entropy of a system of harmonic oscillators in thermal equilibrium. In the
later case $\nu$ represents, of course, the average of the number
operator \cite{a1}.
Eq. (4.2) together with the function $\nu$ defined by (4.3), (4.4)
is the desired entropy
for the system. Although the expression (4.2) for the entropy has a well-known
form, the function $\nu$ induces a specific behaviour of the entropy.
It is clear that the time
dependence of the entropy is given by the damping factors $\exp(-4
\lambda t),~ \exp(-2\lambda t) $ and the oscillating function $\exp
2i\Omega t.$ The complex
oscillating factor $\exp 2i\Omega t$ reduces to a function of the
frequency $\omega,$
namely $\exp 2i\omega t$ for $\mu\to 0$ or if $\mu/\Omega\ll 1$ (i.e.
the frequency $\omega$ is very large as compared to $\mu).$

In the case of a thermal bath (2.7), (2.8), a time-dependent effective
temperature $T_e$ can be defined \cite{lw,j1},
by noticing that when $t\to\infty,$
$\nu$ tends, according to (2.24) in ref. \cite{a}, to the average thermal
phonon number $<n>=(\exp(\hbar\omega/kT)-1)^{-1}.$ Thus $\nu$ can be
considered as giving the time evolution of the thermal phonon number,
so that we can put in this case
\bea  (\exp{\hbar\omega\over kT_e}-1)^{-1}=\nu.\eea
The function $\nu$ vanishes at
$t=0.$ From (4.5) the effective temperature $T_e$ can be expressed as
\bea  T_e(t)={\hbar\omega\over k[\ln(\nu+1)-\ln\nu]}.\eea
Accordingly, we can say that at
time $t$ the system is in thermal equilibrium at temperature $T_e.$
In terms of the effective temperature, the von Neumann entropy takes the form
\bea S(t)={\hbar\omega\over T_e(\exp{\displaystyle {\hbar\omega\over kT_e}}
-1)}-k\ln[1-\exp(-{\hbar\omega\over kT_e})].\eea
As $t$ increases, the
effective temperature approaches thermal equilibrium with the bath, $T_e\to T.$

\section{Concluding remarks}

Recently we assist to a revival of interest in quantum brownian motion as a
paradigm of quantum open systems. There are many motivations. The possibility
of preparing systems in macroscopic quantum states led to the problems of
dissipation in tunneling and of loss of quantum coherence (decoherence). These
problems are intimately related to the issue of quantum-to-classical
transition.
All of them point the necessity of a better understanding of open quantum
systems and require the extension of the model of quantum brownian motion.
Our results allow such extensions.
The Lindblad theory provides a selfconsistent
treatment of damping as a possible extension of quantum mechanics to open
systems. In the present paper we have studied the one-dimensional harmonic
oscillator with dissipation within the framework of this theory.
We have first obtained the explicit form of the density operator from the
master and Fokker-Planck equations. The density operator
in a Gaussian form is a function of the position and momentum operators
in addition to several time dependent
factors. Then the density operator has been used
to calculate the von Neumann entropy and the effective temperature.
The temporal behaviour of these quantities shows how they approach
their equilibrium values.
In a future work we plan to discuss the von Neumann entropy in association
with uncertainty, decoherence and correlations of the system
with its environment \cite{paz1,paz2}.


\begin{thebibliography}{99}

\bibitem{h}
R. W. Hasse, J. Math. Phys. {\bf 16}, 2005 (1975)

\bibitem{d}
E. B. Davies, Quantum Theory of Open Systems (Academic Press, New York, 1976)

\bibitem{s}
H. Spohn, Rev. Mod. Phys. {\bf 52}, 569 (1980)

\bibitem{d2}
H. Dekker, Phys. Rep. {\bf 80}, 1 (1981)

\bibitem{li}
K. H. Li, Phys. Rep. {\bf 134}, 1 (1986)

\bibitem{rev}
A. Isar, A. Sandulescu, H. Scutaru, E. Stefanescu and W. Scheid, Int. J.
Mod. Phys. E {\bf 3}, 635 (1994)

\bibitem{d1}
H. Dekker, Physica A {\bf 95}, 311 (1979)

\bibitem{l1}
G. Lindblad, Commun. Math. Phys. {\bf 48}, 119 (1976)

\bibitem{l2}
G. Lindblad, Rep. Math. Phys. {\bf 10}, 393 (1976)

\bibitem{ss}
A. Sandulescu and H. Scutaru, Ann. Phys. (N.Y.) {\bf 173}, 277 (1987)

\bibitem{i1}
A. Isar, A. Sandulescu and W. Scheid, J. Phys. G - Nucl. Part. Phys. {\bf 17},
385 (1991)

\bibitem{i2}
A. Isar, W. Scheid and A. Sandulescu, J. Math. Phys. {\bf 32}, 2128 (1991)

\bibitem{a}
A. Isar, Helv. Phys. Acta {\bf 67}, 436 (1994)

\bibitem{ass}
A. Isar, A. Sandulescu and W. Scheid, J. Math. Phys. {\bf 34}, 3887 (1993)

\bibitem{hh}
H.Haken, Rev. Mod. Phys. {\bf 47}, 67 (1975)

\bibitem{hr}
H. Risken, The Fokker-Planck Equation (Springer, Berlin, 1984)

\bibitem{lw}
W. H. Louisell and R. L. Walker, Phys. Rev. {\bf 137}, B 204 (1965)

\bibitem{rh}
H. S. Robertson and M. A. Huerto, Phys. Rev. Lett. {\bf 23}, 825 (1969)

\bibitem{a1}
G. S. Agarwal, Phys. Rev. A {\bf 3}, 828 (1971)

\bibitem{m}
J. E. Moyal, Proc. Cambridge Philos. Soc. {\bf 45}, 99 (1949)

\bibitem{j1}
S. Jang, Physica A {\bf 175}, 420 (1991)

\bibitem{wu}
M. C. Wang and G. E. Uhlenbeck, Rev. Mod. Phys. {\bf 17}, 323 (1945)

\bibitem{w}
R. M. Wilcox, J. Math. Phys. {\bf 8}, 962 (1967)

\bibitem{l}
W. H. Louisell, Quantum Statistical Properties of Radiation (Wiley, New York,
1973)

\bibitem{hw}
H. Weyl, The Theory of Group and Quantum Mechanics, (Dover, New York, 1950)

\bibitem{clm}
C. L. Mehta, J. Math. Phys. {\bf 5}, 677 (1964)

\bibitem{j2}
S. Jang, Nucl. Phys. A {\bf 499}, 250 (1989)

\bibitem{paz1}
J. P. Paz, S. Habib and W. Zurek, Phys. Rev. D {\bf 47}, 488 (1993)

\bibitem{paz2}
W. Zurek, S. Habib and J. P. Paz, Phys. Rev. Lett. {\bf 70}, 1187 (1993)

\end{thebibliography}
\end{document}